\documentclass[preprint,aps,amsmath,showpacs,showkeys,nofootinbib, 
superscriptaddress]{revtex4} 
\begin{document}

\title{\textbf{Deeply bound $\Xi$ tribaryon}}
\author{H.~Garcilazo} 
\email{humberto@esfm.ipn.mx} 
\affiliation{Escuela Superior de F\' \i sica y Matem\'aticas, \\ 
Instituto Polit\'ecnico Nacional, Edificio 9, 
07738 M\'exico D.F., Mexico} 

\author{A.~Valcarce} 
\email{valcarce@usal.es} 
\affiliation{Departamento de F\'\i sica Fundamental and IUFFyM,\\ 
Universidad de Salamanca, E-37008 Salamanca, Spain}

\date{\today} 

\begin{abstract}
We have used realistic local interactions based on the recent update 
of the strangeness $-2$ Nijmegen ESC08c potential to calculate the bound state problem of the
$\Xi NN$ system in the $(I)J^P=(\frac{1}{2})\frac{3}{2}^+$ state. 
We found that this system presents a deeply bound state lying 
$13.5$ MeV below the $\Xi d$ threshold. Since in lowest order, 
pure S$-$wave configuration, this system can not decay into 
the open $\Lambda\Lambda N$ channel, its decay width is expected 
to be very small. We have also recalculated the
$(I)J^P=(\frac{3}{2})\frac{1}{2}^+$ state and we 
have compared with results of quark-model based potentials.
\end{abstract}

\maketitle 

\section{Introduction}
The interaction between baryons in the strangeness $-2$ sector has been in 
the focus of interest for many years. The new hybrid experiment $E07$ recently
approved at J--PARC is expected to record of the order of $10^4$ $\Xi^-$ stopping 
events~\cite{Nak15}, one order of magnitude larger than the previous $E373$
experiment. Under the development of an overall scanning method a first output
of this ambitious project was already obtained, the so--called KISO event, the first 
clear evidence of a deeply bound state of $\Xi^{-} - ^{14}$N~\cite{Naa15}.
Together with other indications of certain emulsion data, these findings suggest 
that the average $\Xi N$ interaction may be attractive~\cite{Rij13,Nag15,Rij16}. 
In particular, the ESC08c Nijmegen potential for baryon--baryon 
channels with total strangeness $-2$ predicted an important 
attraction in the isospin 1 $\Xi N$ interaction, with a bound state 
of 8.3 MeV in the $\Xi N$ channel with isospin-spin quantum numbers, $(i,j)=(1,1)$~\cite{Rij13}.
The recent update of the ESC08c Nijmegen potential to take into account 
the new experimental information of Ref.~\cite{Naa15} concludes 
the existence of a bound state in the  
$(i,j)=(1,1)$ $\Xi N$ channel with a binding energy of 1.56 MeV~\cite{Nag15}.
It is worth to mention that the latest, although still preliminary, results
from lattice QCD simulations of the strangeness $-2$ baryon--baryon interactions
also suggest an overall attractive $\Xi N$ interaction~\cite{Sas15}.

In a recent work~\cite{Gar15} we have used the results of
the new ESC08c Nijmegen strangeness $-2$ baryon--baryon 
interaction~\cite{Nag15} to analyze the possible existence 
of $\Xi NN$ bound states in isospin $3/2$ channels.
Our main motivation was the decoupling of $I=3/2$ $\Xi NN$ channels
from the lowest $\Lambda\Lambda N$ channel, due to isospin conservation,
what would make a possible bound state stable. 
In particular, for the case of the $J^P=\frac{1}{2}^+$ state
we showed that the existence of the deuteron--like $\Xi N$ $(i,j^p)=(1,1^+)$ $D^\ast$ 
bound state predicted by the ESC08c Nijmegen model
is sufficient to guarantee the existence of a $\Xi NN$
$J^P=\frac{1}{2}^+$ bound state with a binding
energy of about 2.5 MeV~\cite{Gar15}. Besides, for the 
$J^P=\frac{3}{2}^+$ channel we pointed out that a bound state 
can not exist as a consequence of the Pauli principle,
since this would require two nucleons in a state with total
spin 1 and total isospin 1, which is forbidden in S$-$wave.
The $(I)J^P=(\frac{1}{2})\frac{1}{2}^+$ state has been recently analyzed
making use of the new ESC08 Nijmegen baryon--baryon interactions for the systems 
with strangeness $0$, $-1$, and $-2$, concluding the 
existence of a tribaryon with mass of 3194 MeV, just 
below the $\Xi d$ threshold~\cite{Gar16}.

In this paper we study the existence of deeply bound
states in other partial waves, as a possible guide to future experiments 
at J--PARC. For this purpose we will construct realistic local 
potentials of the ESC08c Nijmegen S$-$wave $\Xi N$ $(i,j) = (0,1)$ and $(1,1)$
interactions, which contribute to the $\Xi NN$ $(I)J^P=(\frac{1}{2})\frac{3}{2}^+$ state. 
We will also construct a realistic local potential of the $\Xi N$ $(i,j) = (1,0)$ channel 
in order to compare with the $(I)J^P=(\frac{3}{2})\frac{1}{2}^+$ $\Xi NN$ 
state which was already studied in our previous work~\cite{Gar15}, but however
considering only the $\Xi N$ $(i,j) = (1,1)$ channel.

In addition to the ESC08c Nijmegen model of the $\Xi N$ interaction we will also consider
the chiral constituent quark model of Ref.~\cite{Car12}. The reason behind this choice
is that these two models present at least one $\Xi N$ bound state. In the case of
the ESC08c model they have incorporated in their analysis the Nagara
event~\cite{Tak01} and the KISO event~\cite{Naa15}. As mentioned above, 
preliminary results from lattice QCD also suggest an overall 
attractive $\Xi N$ interaction~\cite{Sas15}. There are other models for the 
$\Xi N$ interaction, like the hybrid quark--model based analysis of Ref.~\cite{Fuj07},
the effective field theory approach of Ref.~\cite{Hai16}, or even some of the
earlier models of the Nijmegen group~\cite{Sto99} that do not present $\Xi N$
bound states and, in general, the interactions are weakly attractive or repulsive.
Thus, one does expect that these models will give rise to $\Xi NN$ bound states.

The paper is organized as follows. We will use Sec.~\ref{secII} for describing
all technical details to solve the $\Xi NN$ bound--state Faddeev equations. 
In Sec.~\ref{secIII} we will construct the two--body amplitudes needed for the
solution of the bound state three--body problem. Our results will be 
presented and discussed in Sec.~\ref{secIV}. We will also compare with results 
from quark-model based potentials.
Finally, in Sec.~\ref{secV} we summarize our main conclusions.

\section{The $\Xi NN$ bound-state Faddeev equations}
\label{secII}

We will restrict ourselves to the configurations where all three 
particles are in S$-$wave states and assume that $\Xi$ is particle
1 and the two nucleons are particles 2 and 3, so that the Faddeev equations 
for the bound--state problem in the case of three baryons with total
isospin $I$ and total spin $J$ are,
\begin{eqnarray}
T_{i;IJ}^{i_ij_i}(p_iq_i) = &&\sum_{j\ne i}\sum_{i_jj_j}
h_{ij;IJ}^{i_ij_i;i_jj_j}\frac{1}{2}\int_0^\infty q_j^2dq_j \nonumber \\ &&
\times \int_{-1}^1d{\rm cos}\theta\, 
t_{i;i_ij_i}(p_i,p_i^\prime;E-q_i^2/2\nu_i) 
\nonumber \\ &&
\times\frac{1}{E-p_j^2/2\mu_j-q_j^2/2\nu_j}\;
T_{j;IJ}^{i_jj_j}(p_jq_j) \, , 
\label{eq1} 
\end{eqnarray}
where $t_{1;i_1j_1}$ stands for the two--body $NN$ amplitudes
with isospin $i_1$ and spin $j_1$, and  
$t_{2;i_2j_2}$ ($t_{3;i_3j_3}$) for the $\Xi N$ amplitudes
with isospin $i_2$ ($i_3$) and spin $j_2$ ($j_3$).  $p_i$
is the momentum of the pair $jk$ (with $ijk$ an even permutation of
$123$) and $q_i$ the momentum of particle $i$ with respect to the pair
$jk$. $\mu_i$ and $\nu_i$ are the corresponding reduced masses,
\begin{eqnarray}
\mu_i &=& \frac{m_jm_k}{m_j+m_k} \, , \nonumber\\
\nu_i &=& \frac{m_i(m_j+m_k)}{m_i+m_j+m_k} \, ,
\label{eq3}
\end{eqnarray}
and the momenta $p_i^\prime$ and $p_j$ in Eq.~(\ref{eq1}) are given by,
\begin{eqnarray}
p_i^\prime &=& \sqrt{q_j^2+\frac{\mu_i^2}{m_k^2}q_i^2+2\frac{\mu_i}{m_k}
q_iq_j{\rm cos}\theta} \, , \nonumber \\
p_j &=& \sqrt{q_i^2+\frac{\mu_j^2}{m_k^2}q_j^2+2\frac{\mu_j}{m_k}
q_iq_j{\rm cos}\theta} \, .
\label{eq5}
\end{eqnarray}
$h_{ij;IJ}^{i_ij_i;i_jj_j}$ are the spin--isospin coefficients,
\begin{eqnarray}
h_{ij;IJ}^{i_ij_i;i_jj_j}= &&
(-)^{i_j+\tau_j-I}\sqrt{(2i_i+1)(2i_j+1)}
W(\tau_j\tau_kI\tau_i;i_ii_j)
\nonumber \\ && \times
(-)^{j_j+\sigma_j-J}\sqrt{(2j_i+1)(2j_j+1)}
W(\sigma_j\sigma_kJ\sigma_i;j_ij_j) \, , 
\label{eq6}
\end{eqnarray}
where $W$ is the Racah coefficient and $\tau_i$, $i_i$, and $I$ 
($\sigma_i$, $j_i$, and $J$) are the isospins (spins) of particle $i$,
of the pair $jk$, and of the three--body system.

Since the variable $p_i$ in Eq.~(\ref{eq1}) runs from 0 to $\infty$,
it is convenient to make the transformation
\begin{equation}
x_i=\frac{p_i-b}{p_i+b} \, ,
\label{eq7}
\end{equation}
where the new variable $x_i$ runs from $-1$ to $1$ and $b$ is a scale
parameter that has no effect on the solution. With this transformation
Eq.~(\ref{eq1}) takes the form,
\begin{eqnarray}
T_{i;IJ}^{i_ij_i}(x_iq_i) = &&\sum_{j\ne i}\sum_{i_jj_j}
h_{ij;IJ}^{i_ij_i;i_jj_j}\frac{1}{2}\int_0^\infty q_j^2dq_j \nonumber \\ &&
\times \int_{-1}^1d{\rm cos}\theta\; 
t_{i;i_ij_i}(x_i,x_i^\prime;E-q_i^2/2\nu_i) 
\nonumber \\ &&
\times\frac{1}{E-p_j^2/2\mu_j-q_j^2/2\nu_j}\;
T_{j;IJ}^{i_jj_j}(x_jq_j) \, . 
\label{eq8} 
\end{eqnarray}
Since in the amplitude $t_{i;i_ij_i}(x_i,x_i^\prime;e)$ the variables
$x_i$ and $x_i^\prime$ run from $-1$ to $1$, one can expand this amplitude
in terms of Legendre polynomials as,
\begin{equation}
t_{i;i_ij_i}(x_i,x_i^\prime;e)=\sum_{nr}P_n(x_i)\tau_{i;i_ij_i}^{nr}(e)P_r(x'_i) \, ,
\label{eq9}
\end{equation}
where the expansion coefficients are given by,
\begin{equation}
\tau_{i;i_ij_i}^{nr}(e)= \frac{2n+1}{2}\frac{2r+1}{2}\int_{-1}^1dx_i
\int_{-1}^1 dx_i^\prime\; P_n(x_i) 
t_{i;i_ij_i}(x_i,x_i^\prime;e)P_r(x_i^\prime) \, .
\label{eq10} 
\end{equation}
Applying expansion~(\ref{eq9}) in Eq.~(\ref{eq8}) one gets,
\begin{equation}
T_{i;IJ}^{i_ij_i}(x_iq_i) = \sum_n P_n(x_i) T_{i;IJ}^{ni_ij_i}(q_i) \, ,
\label{eq11}
\end{equation}
where $T_{i;IJ}^{ni_ij_i}(q_i)$ satisfies the one--dimensional integral equation,
\begin{equation}
T_{i;IJ}^{ni_ij_i}(q_i) = \sum_{j\ne i}\sum_{mi_jj_j}
\int_0^\infty dq_j A_{ij;IJ}^{ni_ij_i;mi_jj_j}(q_i,q_j;E)\;
T_{j;IJ}^{mi_jj_j}(q_j) \, , 
\label{eq12}
\end{equation}
with
\begin{eqnarray}
A_{ij;IJ}^{ni_ij_i;mi_jj_j}(q_i,q_j;E)= &&
h_{ij;IJ}^{i_ij_i;i_jj_j}\sum_r\tau_{i;i_ij_i}^{nr}(E-q_i^2/2\nu_i)
\frac{q_j^2}{2}
\nonumber \\ &&
\times\int_{-1}^1 d{\rm cos}\theta\;\frac{P_r(x_i^\prime)P_m(x_j)} 
{E-p_j^2/2\mu_j-q_j^2/2\nu_j} \, .
\label{eq13} 
\end{eqnarray}

The three amplitudes $T_{1;IJ}^{ri_1j_1}(q_1)$, $T_{2;IJ}^{mi_2j_2}(q_2)$,
and $T_{3;IJ}^{ni_3j_3}(q_3)$ in Eq.~(\ref{eq12}) are coupled together.
The number of coupled equations can be reduced, however, since two of
the particles are identical. The reduction procedure for the case where
one has two identical fermions has been described before~\cite{Afn74,Gar90}
and will not be repeated here. With the assumption that particle 1 is
the $\Xi$ and particles 2 and 3 are the nucleons, only the amplitudes 
$T_{1;IJ}^{ri_1j_1}(q_1)$ and $T_{2;IJ}^{mi_2j_2}(q_2)$ are independent
from each other and they satisfy the coupled integral equations,
\begin{equation}
T_{1;IJ}^{ri_1j_1}(q_1)  =  2\sum_{mi_2j_2}
\int_0^\infty dq_3 A_{13;IJ}^{ri_1j_1;mi_2j_2}(q_1,q_3;E)\;
T_{2;IJ}^{mi_2j_2}(q_3) \, ,
\label{eq14} 
\end{equation}
\begin{eqnarray}
T_{2;IJ}^{ni_2j_2}(q_2) = && \sum_{mi_3j_3}g
\int_0^\infty dq_3 A_{23;IJ}^{ni_2j_2;mi_3j_3}(q_2,q_3;E)\;
T_{2;IJ}^{mi_3j_3}(q_3) 
\nonumber \\ && +
\sum_{ri_1j_1}
\int_0^\infty dq_1 A_{31;IJ}^{ni_2j_2;ri_1j_1}(q_2,q_1;E)\;
T_{1;IJ}^{ri_1j_1}(q_1) \, , 
\label{eq15} 
\end{eqnarray}
with the identical--particle factor
\begin{equation}
g=(-)^{1+\sigma_1+\sigma3-j_2+\tau_1+\tau_3-i_2} \, ,
\label{eq16}
\end{equation}
where $\sigma_1$ ($\tau_1$) and $\sigma_3$ ($\tau_3$) 
stand for the spin (isospin) of the $\Xi$ and the $N$, respectively.

Substitution of Eq.~(\ref{eq14}) into Eq.~(\ref{eq15}) yields an
equation with only the amplitude $T_2$,
\begin{equation}
T_{2;IJ}^{ni_2j_2}(q_2) = \sum_{mi_3j_3}
\int_0^\infty dq_3 K_{IJ}^{ni_2j_2;mi_3j_3}(q_2,q_3;E)\;
T_{2;IJ}^{mi_3j_3}(q_3) \, , 
\label{eq17}
\end{equation}
where
\begin{eqnarray}
K_{IJ}^{ni_2j_2;mi_3j_3}(q_2,q_3;E)= && g
A_{23;IJ}^{ni_2j_2;mi_3j_3}(q_2,q_3;E)
\nonumber \\ && +
2\sum_{ri_1j_1}
\int_0^\infty dq_1 A_{31;IJ}^{ni_2j_2;ri_1j_1}(q_2,q_1;E)
\nonumber \\ && \times
A_{13;IJ}^{ri_1j_1;mi_3j_3}(q_1,q_3;E) \, .
\label{eq18} 
\end{eqnarray}

\section{Two--body amplitudes}
\label{secIII}

We have constructed the two--body amplitudes by solving the Lippmann--Schwinger
equation of each $(i,j)$ channel,
\begin{eqnarray}
t^{ij}(p,p';e)= && V^{ij}(p,p')+\int_0^\infty {p^{\prime\prime}}^2
dp^{\prime\prime} V^{ij}(p,p^{\prime\prime})
\nonumber \\ && \times
\frac{1}{e-{p^{\prime\prime}}^2/2\mu} t^{ij}(p^{\prime\prime},p';e) \, ,
\label{eq19} 
\end{eqnarray}
where 
\begin{equation}
V^{ij}(p,p')=\frac{2}{\pi}\int_0^\infty r^2dr\; j_0(pr)V^{ij}(r)j_0(p'r) \, ,
\label{eq20} 
\end{equation}
and the two--body potentials consist of an attractive and a repulsive
Yukawa term, i.e.,
\begin{equation}
V^{ij}(r)=-A\frac{e^{-\mu_Ar}}{r}+B\frac{e^{-\mu_Br}}{r} \, ,
\label{eq21} 
\end{equation}
where the parameters of the three $\Xi N$ channels were obtained by
fitting the low--energy data of each channel of Ref.~\cite{Nag15}. The 
parameters of these models are given in Table~\ref{t1}. In the case 
of the $NN$ $(0,1)$ and $(1,0)$ channels we used
the Malfliet--Tjon models with the parameters given in Ref.~\cite{Gib90}.
\begin{table}[t]
\caption{Low--energy parameters of the $\Xi N$ channels of 
the ESC08c Nijmegen interactions of Ref.~\cite{Nag15}
and the parameters of the corresponding local potentials given
by Eq.~(\ref{eq21}).}
\begin{ruledtabular} 
\begin{tabular}{ccccccccc} 
& $(i,j)$ & $a({\rm fm})$ & $r_0({\rm fm})$ & $A$(MeV fm) & 
$\mu_A({\rm fm}^{-1}$) 
& $B$(MeV fm) & $\mu_B({\rm fm}^{-1})$  & \\
\hline
& $(0,1)$ & $-5.357$ & $1.434$  &  $377$  & $2.68$  & $980$ & $6.61$ & \\ 
& $(1,0)$ & $0.579$  & $-2.521$ &  $290$  & $3.05$  & $155$ & $1.60$ & \\
& $(1,1)$ & $4.911$  & $0.527$  &  $568$  & $4.56$  & $425$ & $6.73$ & \\
\end{tabular}
\end{ruledtabular}
\label{t1} 
\end{table}

\section{Results and discussion}
\label{secIV}

We show in Table~\ref{t2} the results for both the $(I)J^P=(\frac{1}{2})\frac{3}{2}^+$ 
and $(I)J^P=(\frac{3}{2})\frac{1}{2}^+$ $\Xi NN$ states calculated with the 
ESC08c Nijmegen interactions of Table~\ref{t1}, as well as the results obtained
with a quark-model based potential, the chiral constituent quark model (CCQM) 
of Ref.~\cite{Car12}. These binding energies are
measured with respect to the lowest threshold which in the case of the
$(\frac{1}{2})\frac{3}{2}^+$ state is the $\Xi d$ threshold and in the 
case of the $(\frac{3}{2})\frac{1}{2}^+$ state are the $N D^*$ threshold
for the ESC08c Nijmegen model ($1.56$ MeV below the $\Xi NN$ mass) and the 
$N - \Xi N_{(1,0)}$ threshold for the CCQM model ($4.8$ MeV below the $\Xi NN$ mass).

As one can check, the binding energy of the $(\frac{3}{2})\frac{1}{2}^+$ state of the
ESC08c Nijmegen model is smaller than that obtained in Ref.~\cite{Gar15} ($2.50$ MeV) since we 
have now included in addition to the $\Xi N$ $(1,1)$ channel also the $\Xi N$ $(1,0)$ 
channel, which is mainly repulsive. In the case of the CCQM, in Ref.~\cite{Gar15} we
had performed the calculation considering both $\Xi N$ channels, thus in perfect agreement
with the present results. Let us note again that the binding energies obtained from 
the CCQM are much smaller than those obtained from the ESC08c Nijmegen model, since 
for the $(I)J^P=(\frac{3}{2})\frac{1}{2}^+$ state the dominant channel is the
$(i,j)=(1,1)$ $N\Xi$ subsystem, that it is almost bound with the CCQM model while 
it has a binding energy of $1.56$ MeV with the ESC08c Nijmegen model~\cite{Nag15}.

The most interesting result of Table~\ref{t2} is the very large binding
energy of the $(\frac{1}{2})\frac{3}{2}^+$ state predicted by the
ESC08c Nijmegen potential model, which would make it easy to identify experimentally as a sharp resonance 
lying some $15.7$ MeV below the $\Xi NN$ threshold.
The $\Lambda\Lambda - \Xi N$ $(i,j)=(0,0)$ transition channel,
which is responsible for the decay $\Xi NN\to\Lambda\Lambda N$,
does not contribute to the 
$(I)J^P=(\frac{1}{2})\frac{3}{2}^+$ state in a pure S$-$wave configuration.
One would need at least the spectator nucleon to be in a D wave or that the
$\Lambda\Lambda - \Xi N$ transition channel be in one of the 
negative parity $P$-wave channels with the nucleon spectator also in a P$-$wave, 
so that due to the angular momentum barriers the resulting
decay width is expected to be very small. We finally note that 
the binding energy obtained from the CCQM is larger that in the
$I=3/2$ channel, but once again much smaller than the
prediction of the recent update of the 
ESC08c Nijmegen potential~\cite{Nag15,Rij16} incorporating the recent experimental
results of Ref.~\cite{Nak15}.
\begin{table}[t]
\caption{Binding energies of the $\Xi NN$ 
$(I)J^P=(\frac{1}{2})\frac{3}{2}^+$ and $(I)J^P=(\frac{3}{2})\frac{1}{2}^+$
states (in MeV) calculated with the interactions based in the 
ESC08c Nijmegen model~\cite{Nag15} and with those of a chiral
constituent quark model~\cite{Car12}.} 
\begin{ruledtabular} 
\begin{tabular}{ccccc} 
& Model & $(\frac{1}{2})\frac{3}{2}^+$ & $(\frac{3}{2})\frac{1}{2}^+$ & \\
\hline
& ESC08c & 13.54 & 1.33   & \\
& CCQM   & 1.15  & 0.43   & \\
\end{tabular}
\end{ruledtabular}
\label{t2} 
\end{table}

\section{Summary}
\label{secV}
Recent results in the strangeness $-2$ sector, the so-called KISO event,
reported clear evidence of a deeply bound state of $\Xi^{-} - ^{14}$N
what could point out that the average $\Xi N$ interaction might be attractive. 
We have made use of the recent update of the ESC08c Nijmegen potential taking into account 
the recent experimental information, to study the bound state problem of the
$\Xi NN$ system in the $(I)J^P=(\frac{1}{2})\frac{3}{2}^+$ state. 
We have found that this system has a deeply bound state that lies 
13.5 MeV below the $\Xi d$ threshold. Since at lowest order, 
pure S$-$wave configuration, this system can not decay into 
the open $\Lambda\Lambda N$ channel, its decay width is expected 
to be very small, what would make it very easy to be identified experimentally. 

The huge amount of $\Xi^-$ stopping events that will be recorded
at the recently approved hybrid experiment $E07$ at J--PARC, 
is expected to shed light on the uncertainties of our knowledge of the baryon-baryon
interaction in the strangeness $-2$ sector.
Meanwhile the scarce experimental information gives rise to an ample room
for speculation.
The present detailed theoretical investigation of the possible existence 
of deeply bound states based on realistic models are basic tools to 
advance in the knowledge of the details of the $\Xi N$ interaction.
First, it could help to raise the awareness of the experimentalist 
that it is worthwhile to investigate few-baryon systems, specifically 
because for some quantum numbers such states could be stable. Secondly, 
it makes clear that strong and attractive $\Xi N$ interactions, like 
those suggested by the ESC08c Nijmegen model,
have consequences for the few-body sector and can be easily tested
against future data.
Observations like the ones reported in Ref.~\cite{Naa15}
are interesting. However, in this case microscopic calculations are
impossible and, consequently, their interpretation will be always
afflicted by large uncertainties.
The identification of strangeness $-2$ hypernuclei in
coming experiments at J--PARC would contribute significantly to
understand nuclear structure and baryon--baryon interactions in 
the strangeness $-2$ sector.

\acknowledgments 
This work has been partially funded by COFAA-IPN (M\'exico) and 
by Ministerio de Educaci\'on y Ciencia and EU FEDER under 
Contracts No. FPA2013-47443-C2-2-P and FPA2015-69714-REDT.

\end{document}